\documentclass[lettersize,journal]{IEEEtran}
\usepackage{amsmath,amsfonts}
\usepackage{array}
\usepackage[caption=false,font=normalsize,labelfont=sf,textfont=sf]{subfig}

\usepackage{textcomp}
\usepackage{bm}
\usepackage{stfloats}
\usepackage{url}
\usepackage{verbatim}
\usepackage{graphicx}
\usepackage{cite}
\usepackage{makecell}
\usepackage{multicol,multirow}
\usepackage[linesnumbered,lined,ruled,commentsnumbered]{algorithm2e}
\usepackage{diagbox} 
\usepackage{booktabs} 
\usepackage{pifont}
\usepackage[normalem]{ulem}
\usepackage{amsmath}
\usepackage{enumitem}
\usepackage{url,hyperref,cleveref}

\usepackage{mathtools}

\usepackage{xcolor}

\newcommand{\black}[1]{\textcolor{black}{#1}}

\begin{document}





\title{MFA-KWS: Effective Keyword Spotting with Multi-head Frame-asynchronous Decoding}

\author{
Yu Xi,~\IEEEmembership{Student Member,~IEEE},
Haoyu Li,~\IEEEmembership{Student Member,~IEEE},
Xiaoyu Gu,\\
Yidi Jiang,~\IEEEmembership{Student Member,~IEEE} and
Kai Yu,~\IEEEmembership{Senior Member,~IEEE}

\thanks{This work was supported in part by China NSFC Project under Grant 92370206, in part by Shanghai Municipal Science and Technology Major Project under Grant 2021SHZDZX0102, and in part by Key Research and Development Program of Jiangsu Province, China, under Grant BE2022059.

Yu Xi, Haoyu Li, Xiaoyu Gu and Kai Yu are with the X-LANCE Lab, School of Computer Science \& MoE Key Lab of Artificial Intelligence, Shanghai Jiao Tong University, Shanghai, 200240, P.~R.~China (\{yuxi.cs, haoyu.li.cs, kai.yu\}@sjtu.edu.cn). They are also with Jiangsu Key Lab of Language Computing, Suzhou, 215021, P.~R.~China.~\textit{(Corresponding author: Kai Yu.)}

Yidi Jiang is with the Department of Electrical and Computer Engineering, National University of Singapore, Singapore 117583~(yidi\_jiang@u.nus.edu).
}
}




\maketitle

\begin{abstract}

Keyword spotting (KWS) is essential for voice-driven applications, demanding both accuracy and efficiency. Traditional ASR-based KWS methods, such as greedy and beam search, explore the entire search space without explicitly prioritizing keyword detection, often leading to suboptimal performance. In this paper, we propose an effective keyword-specific KWS framework by introducing a streaming-oriented CTC-Transducer-combined frame-asynchronous system with multi-head frame-asynchronous decoding (MFA-KWS). Specifically, MFA-KWS employs keyword-specific phone-synchronous decoding for CTC and replaces conventional RNN-T with Token-and-Duration Transducer to enhance both performance and efficiency. Furthermore, we explore various score fusion strategies, including single-frame-based and consistency-based methods. Extensive experiments demonstrate the superior performance of MFA-KWS, which achieves state-of-the-art results on both fixed keyword and arbitrary keywords datasets, such as Snips, MobvoiHotwords, and LibriKWS-20, while exhibiting strong robustness in noisy environments. Among fusion strategies, the consistency-based CDC-Last method delivers the best performance. Additionally, MFA-KWS achieves a 47\%–63\% speed-up over the frame-synchronous baselines across various datasets, with a lightweight model size of approximately 3.3M parameters, making it well-suited for on-device deployment.

\end{abstract}



\section{Introduction}

\IEEEPARstart{K}{eyword} spotting (KWS)~\cite{hoy2018alexa,ieee_access-lopez-kws_overview}, emphasized as \textit{wake word detection} (WWD) aims to detect the pre-defined keywords from continuous streaming audio in memory- and computation-constrained environments, serving as the main interactive entrance for intelligent assistants. With the rapid advancements in multi-modal large language models and end-to-end speech foundation models~\cite{emnlpfinding2023-speechgpt,arxiv2024-pengjing-slmsurvey,arxiv2024-llamaomni,arxiv2024-omniflatten,arxiv2024-moshi,arxiv2024-glm4voice}, creating more intelligent and personalized assistants has become increasingly feasible, drawing significant attention from users. In this context, designing a powerful and efficient interactive interface is becoming increasingly critical.

KWS systems can generally be roughly classified into two categories. The first category involves pattern matching~\cite{icassp2015-new-qbye-1,interspeech2017-new-qbye-2,arxiv2018-new-qbye-3,interspeech2019-qbye-at-01,icassp2021-new-qbye-4,icassp2021-new-qbye-5,interspeech2022-new-qbye-6,interspeech2022-qbye-at-02-libriphrase,icassp2023-qbye_01,icassp2023-qbye_02,interspeech2023-qbye-at-03-libriphrase-apple,interspeech2023-qbye-at-04-libriphrase-korea,icassp2024-yuxi-cobe,icassp2024-apple-openvocab-at} or multi-label classification~\cite{icassp2014-guoguochen-dnn_kws,icassp2015-automatic-gain-control,icassp2019-Coucke-second_baseline_of_snips_in_wekws,is2020-kunzhang-first_baseline_of_snips_in_wekws,interspeech2020-google-end-to-end,icassp2023-jiewang-wekws,icassp2024-haizhu-tcsnkws}, where KWS models process audio chunks (or frames) in a streaming mode, determining whether each chunk is activated for a specific keyword or identifying which keyword it belongs to. This approach is straightforward and follows an end-to-end manner. However, as a segment-level classification model, the system is data-sensitive. While it may perform well in quiet or controlled environments, its performance is inconsistent across diverse scenarios and it lacks robustness to noise. 

Another widely adopted approach leverages automatic speech recognition (ASR) within a two-stage framework. ASR criteria are used to train acoustic models~\cite{cnn-kws,lstm-kws-1,lstm-kws-2,att-kws-1,att-kws-2}, while decoding algorithms~\cite{paper-decoding-1,icassp2021-zuozhen-kws_transducer3,interspeech2022-zhanhengyang-catt-kws,asru2023-aozhang-u2kws_ctckws2,icassp2024-yuxi-tdt_kws,icassp2025-yuxi-cdc_kws} process the frame-level posteriors. Compared to classification-based training, sequence-to-sequence ASR training enables acoustic models to adapt to diverse acoustic conditions, ensuring strong performance, particularly in complex environments~\cite{li24r_interspeech,icassp2025-yuxi-ntc_kws,icassp2025-yuxi-cdc_kws}. Moreover, ASR-based KWS supports arbitrary keyword detection~\cite{icassp2022-yuxu-text_adaptive_arbiKWS,icassp2024-yuxi-tdt_kws}, providing greater flexibility for user-defined scenarios. The main challenge with ASR-based systems lies in designing effective and efficient decoding algorithms tailored to KWS task. Traditionally, researchers construct weighted finite-state transducer (WFST)-based decoding graphs~\cite{hmm-filler-1,hmm-filler-2,asru2017-yanzhanghe-classic_ctc_rnnt_kws,hmm-filler-4,icassp2020-sharma-kws_transducer1,interspeech2022-zhanhengyang-catt-kws,icassp2025-yuxi-ntc_kws} that integrate both keyword and filler paths. However, WFST-based decoding is complex to implement and maintain. Alternative methods, such as greedy search or prefix beam search for connectionist temporal classification (CTC)-based KWS~\cite{icassp2023-jiewang-wekws} and greedy or autoregressive beam search for RNN-T-based systems~\cite{icassp2021-tian-kws_transducer2}, explore the full search space without explicitly prioritizing keywords, often leading to suboptimal results. Therefore, developing effective, keyword-specific algorithms for streaming ASR-based KWS remains a promising research direction.


RNN-T~\cite{icml2012-graves-rnnt}, also known as Transducer\footnote{Recurrent networks were originally used as the encoder in the Transducer architecture, which led to the name RNN-T. However, modern Transducers typically use various networks for the encoder. In this work, we use the terms RNN-T to refer to encoder-predictor-joiner model that employs the conventional Transducer loss and use Transducer to refer to any kind of Transducer, such as RNN-T or TDT.}, is a mainstream architecture for ASR~\cite{interspeech2019-tian-rnnt_asr_01,arxiv2019-ching-rnnt_asr_02,icassp2020-qian-rnnt_asr_03,interspeech2020-wei-rnnt_asr_04,Tiny-RNNT} and KWS~\cite{asru2017-yanzhanghe-classic_ctc_rnnt_kws,icassp2020-sharma-kws_transducer1,icassp2021-tian-kws_transducer2,icassp2021-zuozhen-kws_transducer3,icassp2024-yuxi-tdt_kws} in both academia and industry. Its inherent support for streaming inference makes it well-suited for KWS, where real-time feedback is essential. In our previous conference work, we proposed TDT-KWS~\cite{icassp2024-yuxi-tdt_kws}, a Transducer-based system that achieves faster and more accurate performance than the original RNN-T-based system. This is achieved through two key modifications: (1) replacing the standard RNN-T with the Token-and-Duration Transducer (TDT), which simultaneously predicts tokens and their durations, and (2) designing a streaming decoding algorithm specifically for Transducer-based KWS. Although TDT-KWS significantly outperforms strong Transducer-based baselines, it can be still improved. Since posterior predictions for the current frame depend on the history of the predicted sequence, error accumulation can occur. The dependency on previous predictions may hinder the ability of the model to distinguish between positive and challenging negative samples, leading to performance degradation in complex acoustic conditions. Ensuring robust performance under challenging scenarios, such as noisy environments or arbitrary keyword detection, still requires further investigation.

Joint multi-task training is a promising direction for addressing this challenge in KWS. By enabling mutual learning and multi-stage detection across tasks, multi-task training enhances robustness and generalization. Moreover, ASR-derived systems offer more powerful and flexible approaches for arbitrary keyword detection. Recent studies on joint training frameworks such as CTC with AED (CTC-AED) \cite{icassp2017-suyoun-aed_ctc_asr,acl2017-takaaki-aed_ctc_joint_decoding,journal2017-shinji-hybrid_aed_ctc,interspeech2018-shinji-espnet} or CTC with Transducer (CTC-Transducer)~\cite{nemo} have demonstrated superior performance over single-architecture models and the joint training strategy has become a standard paradigm in ASR. Multi-task systems leverage the strengths of both branches while mitigating individual limitations during training. Specifically, CTC is commonly applied to the acoustic encoder in AED to improve convergence and constrain the attention model from generating overly flexible outputs~\cite{icassp2017-suyoun-aed_ctc_asr}. 
Consequently, multi-head decoding strategies, such as CTC/AED decoding~\cite{acl2017-takaaki-aed_ctc_joint_decoding} were proposed to further enhance transcription accuracy during inference. However, for our KWS goal, joint CTC-AED systems are not universally suitable~\cite{taslp2021-runyanyang-ctc_las_keyword_search,asru2023-aozhang-u2kws_ctckws2} due to the high computational cost of attention mechanisms and the significant latency introduced by token-level inference. Thus, designing an efficient joint framework and multi-head decoding algorithms for KWS remains challenges.

In this paper, we propose \textit{multi-head frame-synchronous decoding} (MFS) to extend our previous Transducer-based KWS system into a joint CTC-Transducer framework. To further improve KWS performance and computational efficiency, we introduce \textit{multi-head frame-asynchronous decoding} (MFA). MFA-KWS exploits the complementary characteristics of CTC and Transducer models to improve single-branch performance, while employing phoneme-level frame-skipping mechanisms to accelerate decoding and reduce computational cost. Compared to a single Transducer-based KWS system, the condition-independent nature of CTC~\cite{icml2006-graves-ctc} mitigates error accumulation, ensuring more stable performance across diverse challenging scenarios. Inspired by phone-synchronous decoding (PSD)~\cite{interspeech2016-zhehuai-psd,taslp2017-zhehuai-psd} in CTC and the Token-and-Duration Transducer (TDT)~\cite{icml2023-hainanxu-tdt} in Transducer-based models, we introduce a KWS-specific PSD for CTC streaming inference and token-duration prediction for Transducer streaming inference. These mechanisms enable frame skipping, improving robustness in noisy conditions while accelerating decoding. 

This work is built upon our previous conference paper, TDT-KWS~\cite{icassp2024-yuxi-tdt_kws}, where we introduced the Token-and-Duration Transducer and a Transducer-based streaming decoding algorithm for KWS. In this study, we extend the Transducer-based approach to a joint CTC-Transducer framework, further improving performance and robustness. The key contributions are as follows: 
\begin{itemize}
    \item We propose MFS-KWS, an effective multi-task KWS system. Building on our previous work, we extend Transducer-based KWS to a CTC-Transducer joint training framework with multi-head frame-synchronous decoding, achieving robust performance across challenging scenarios, including hard negative samples and noisy conditions.
    \item To further enhance performance and accelerate decoding, we propose MFA-KWS, a multi-head frame-asynchronous decoding framework that integrates PSD for the CTC branch and TDT for the Transducer branch. To fully exploit the advantages of MFA-KWS, we design and explore several novel fusion strategies.
    \item We evaluate MFA-KWS on fixed-keyword English and Mandarin datasets (Hey Snips and MobvoiHotwords) and an arbitrary-keyword dataset (LibriKWS-20, derived from LibriSpeech). Results show that MFA-KWS achieves state-of-the-art (SOTA) performance across diverse datasets. Additionally, evaluations across multiple signal-to-noise ratio (SNR) levels demonstrate that MFA-KWS significantly outperforms single-branch systems in noise robustness.
    \item The proposed MFA-KWS achieves a 47\%–63\% inference speed-up across different test sets compared to the frame-synchronous MFS framework. Compared to single-branch KWS frameworks, MFA-KWS strikes a better balance between performance and efficiency, demonstrating its strong potential for on-device KWS applications.
    \item Open-source KWS decoding research remains limited, particularly for KWS decoding algorithms. To advance the development of KWS, we release all decoding codes, including  MFA, Transducer streaming decoding, and other related algorithms (e.g., CTC-specific streaming decoding, multiple fusion strategies) at Github~\footnote{https://github.com/X-LANCE/KWStreamingSearch}.
\end{itemize}

\section{Methodology}
\begin{figure*}[h]
    \centering
    \includegraphics[width=1.0\linewidth]{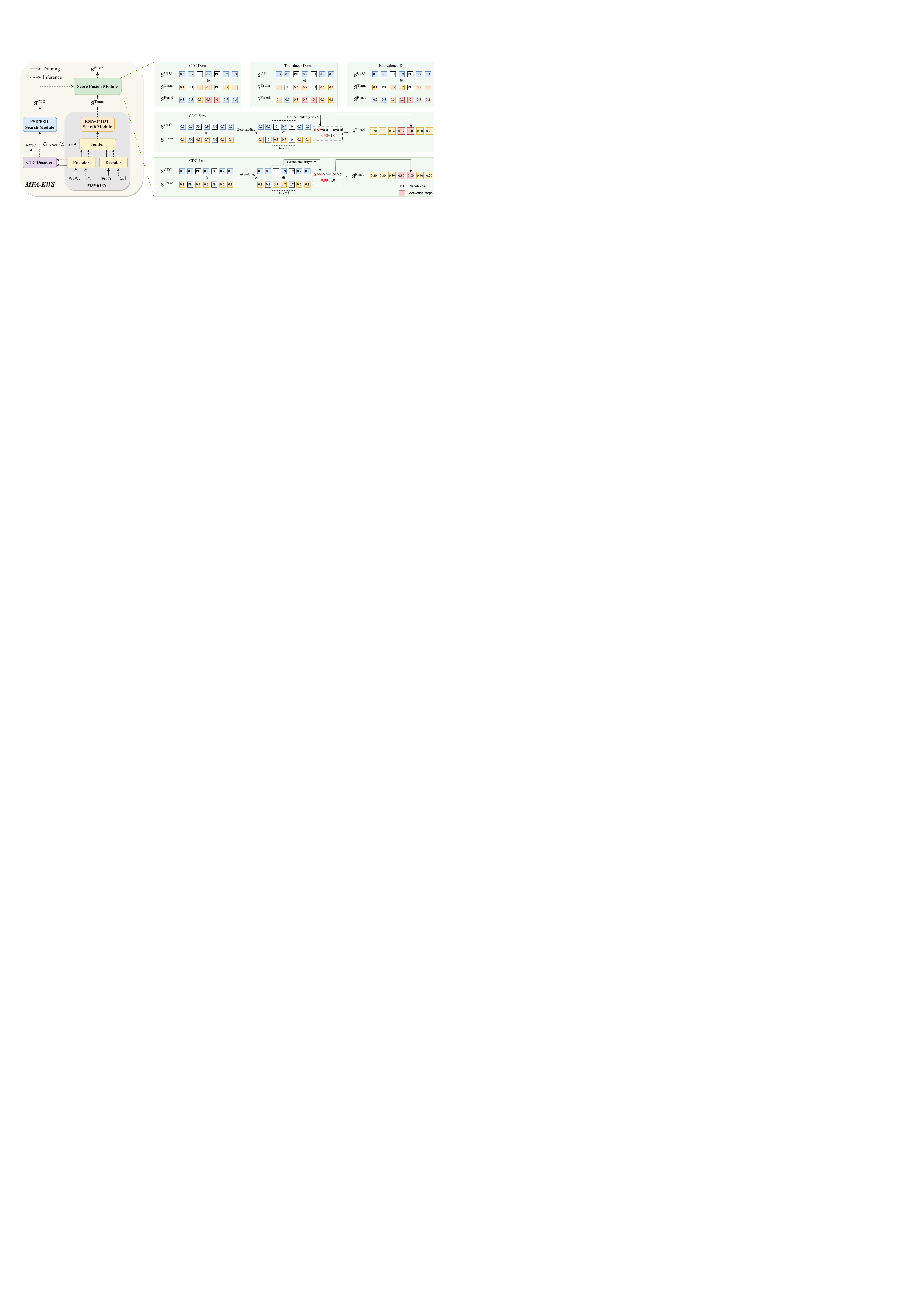}
    \caption{Overview of the MFA-KWS framework. The left part illustrates the MFA-KWS framework, including the training and inference pipelines. The right part presents various fusion strategies for multi-head decoding, as introduced in \Cref{sec:fusion_strategy}. PH~(placeholder) represents a special state for skipped time steps in frame-asynchronous decoding. Blue blocks indicate CTC scores, orange blocks represent Transducer scores and red blocks highlight activation steps.}
    \label{fig:overview}
\end{figure*}

This section provides a comprehensive overview of our work. We first introduce the design of our Transducer-based keyword spotting system, detailing its architecture and decoding algorithm. Next, we propose MFS-KWS, leveraging multi-task learning within the CTC-Transducer joint training framework to enhance TDT-KWS. We also present a streaming multi-head decoding algorithm for MFS-KWS, which merges scores in a frame-synchronous manner. Finally, we extend MFS-KWS to MFA-KWS, enabling more effective and efficient frame-asynchronous decoding.

\begin{algorithm}[t]
\caption{Transducer Streaming Decoding}
 \label{alg:alg_rnnt_tdt_algo}
 \KwIn{Phoneme/blank posterior matrix $\mathbf{p} = [p_1,p_2,\cdots,p_T] \in \mathbb{R}^{T\times U\times V}$, \\ \quad\quad\quad Keyword phoneme sequence \\ \quad\quad\quad $\mathbf{y}=[y_1,y_2,\cdots,y_U]$, \\ \quad\quad\quad Bonus score: $S_{\mathrm{Bonus}}$, \\
 \quad\quad\quad Timeout: $T_{\mathrm{out}}$.}
 \KwOut{Activation scores $\mathbf{S}[1:T]$}
 \BlankLine 
 Init 1): Insert a RNN-T blank token to $\mathbf{y}$ and update $\mathbf{y} =[\phi_{\text{RNN-T}}, y_1,y_2,\cdots,y_U]$ by inserting $\phi_{\text{RNN-T}}$ prior to $\mathbf{y}$. 
 
 Init 2): $\mathbf{S}[1:T] = \{0\}$. 
 
 Init 3): $\phi(0, u) = 0,  \delta(0, u) = 1 \text{ for } 0 \le u \le U$.
 
 Init 4): $t = 1$, $d = 1$, $G = \{\phi_{\text{RNN-T}}\}$. \# \textbf{G stores the greedy results for duration prediction.}

 \While{$t\le T$}{

    \If{ $u = 0 $} {

    $\delta(t, u) = 1$ \# \textbf{new competitor starts at time $t$}

    }
    
    \For{$u = 1$ \KwTo $U$}{


        $\delta(t, u) = \max(\delta(t, u-1) \cdot p_{t, u-1}(\mathbf{y}_{u}),\delta(t-d, u) \cdot p_{t-d, u}(\phi_{\text{RNN-T}})$ \;
        
    }
    $\mathbf{S}[t] = \delta(t, U) \cdot \phi(t, U)$ \;
    
    \uIf{\text{RNN-T KWS}}{
        $d = 1$\;
    }
    \ElseIf{\text{TDT KWS}}{
        $d = \operatorname*{argmax}_{d} P_D(d|t, G)$ \;
        $v = \operatorname*{argmax}_{v} P_T(v|t, G)$ \;
        $G \leftarrow G \cup \{v\}$ \;
        \For{$i\leftarrow 1$ \KwTo $d-1$}{
            $\mathbf{S}[t + i] = 0$\;
        }
    }
    
    \BlankLine
    
    \# \textbf{record the length of the max score path as $\ell(t)$}

    $\ell(t) = \text{RecordPathLength}(\mathbf{S}[t])$

    \BlankLine

    \If{ $\ell(t)> T_{\mathrm{out}}$   } {  
        \# \textbf{discard paths longer than $T_{\mathrm{out}}$} \\
        $\mathbf{S}[t]=0$\; 
    } 

    $\mathbf{S}[t]=\text{pow}(S_{\mathrm{Bonus}} \cdot \mathbf{S}[t],\, 1 / \ell(t))$ \; 

    $t = t + d$ \;
  }
 \Return $\mathbf{S}[1:T]$
\end{algorithm}

\begin{algorithm}[t]
    \caption{Multi-head Streaming Decoding}
    \label{alg:mfs_joint_decoding}
    \KwIn{Transducer posterior matrix $\mathbf{p}^{\text{Trans}} = [p_1^{\text{Trans}},p_2^{\text{Trans}},\cdots,p_T^{\text{Trans}}] \in \mathbb{R}^{T\times U\times V}$, 
    \\ \quad\quad\quad CTC posterior matrix \\ \quad\quad\quad  $\mathbf{p}^{\text{CTC}} = [p^{\text{CTC}}_1,p^{\text{CTC}}_2,\cdots,p^{\text{CTC}}_T] \in \mathbb{R}^{T\times V}$, \\ \quad\quad\quad Keyword phoneme sequence  \\ \quad\quad\quad $\mathbf{y}=[y_1,y_2,\cdots,y_U]$, 
    \\ \quad\quad\quad Bonus score: $S_{\text{bonus}}$,  \\ \quad\quad\quad Timeout: $T_{\text{out}}$ \\ \quad\quad\quad CTC-PSD blank pruning threshold: $\lambda_{\phi}$.} 
    \KwOut{$\mathbf{S}[T]$}  
    \BlankLine

    \SetKw{Or}{or}
    \SetKw{Continue}{continue}
    
    \For{$t=1$ \Or $T$} {

        \# \textbf{return the RNN-T/TDT score at step $t$.}\\
        $\mathbf{S}^{\text{Trans}}$  = Transducer\_Streaming\_Decoding(\\
        \quad $\mathbf{p}^{\text{Trans}}_{[1:t]}$, $\mathbf{y}$, $S_{\text{bonus}}$, $T_{\text{out}}$)
        
        \BlankLine

        \# \textbf{return the CTC score at step $t$.} \\
        $\mathbf{S}^{\text{CTC}}$ = CTC\_Streaming\_Decoding(\\ \quad$\mathbf{p}^{\text{CTC}}_{[1:t]}$, $\mathbf{y}$, $S_{\text{bonus}}$, $T_{\text{out}}$, $\lambda_{\phi}$)
        
        \BlankLine 

        \If{$ \mathbf{S}^{\text{Trans}}[t] = 0 $}{
        $\mathbf{S}^{\text{Trans}}[t]$ = PH  \# \textbf{PH denotes placeholder.} 
        } 

        \If{$ \mathbf{S}^{\text{CTC}}[t] = 0 $}{
        
        
        $\mathbf{S}^{\text{CTC}}[t]$ = PH 
        
        } 
        \# \textbf{Calculated by different fused strategies.} \\
        $\mathbf{S}[t]$ = Score\_Fusion\_Module($\mathbf{S}^{\text{Trans}}$, $\mathbf{S}^{\text{CTC}}$) 
        }
    \Return $\mathbf{S}[1:T]$ 
\end{algorithm}
        
        

        
                
                
                
                

    



    


\subsection{Transducer-based Keyword Spotting System}
\label{sec:2.1}
A Transducer comprises three components: an encoder, a predictor, and a joiner (or joint network). The encoder converts speech signals into latent acoustic representations, while the predictor functions as a language model, processing text input—typically phonemes, graphemes, or sub-words to provide linguistic context auto-regressively. The latent acoustic and textual information from these modules is then merged by the joiner, which generates the probability distribution of the next token. Given an audio input $\mathbf{x} = \{x_{1},x_{2},...,x_{T}\} \in \mathbb{R}^{T \times D}$, where each $x_{t}$ is a $D$-dimensional acoustic feature vector and $T$ denotes the total number of frames, and a transcription $\mathbf{y} = \{y_{1},y_{2},...,y_{U}\} \in \mathbb{R}^{U \times 1}$, where $y_{u}$ represents a label token and $U$ is the total number of tokens, the Transducer is optimized to maximize the log-probability of the ground-truth transcription $\mathbf{y}$ given $\mathbf{x}$. The loss function is formulated as:
\begin{align}
\hspace{-0.5em} \mathcal{L}_\text{RNN-T}(\mathbf{x},\mathbf{y})=-\log p(\mathbf{y}|\mathbf{x})=-\log \,
\sum_{\mathclap{\hspace{1em} \bm{\pi}_\text{RNN-T} \in \mathcal{B}_\text{RNN-T}^{-1}}} \hspace{0.5em} 
p(\bm{\pi}_\text{RNN-T}|\mathbf{x}),
\end{align}
where $\mathcal{B}_\text{RNN-T}$ is defined as a mapping from legal Transducer-based augmented alignments $\bm{\pi}_\text{RNN-T}$~(including a special blank $\phi_{\text{RNN-T}}$) to the label $\mathbf{y}$ and $\mathcal{B}_\text{RNN-T}^{-1}$ refers to the inverse mapping.

\begin{figure}[h]
  \centering
     \includegraphics[width=\linewidth]{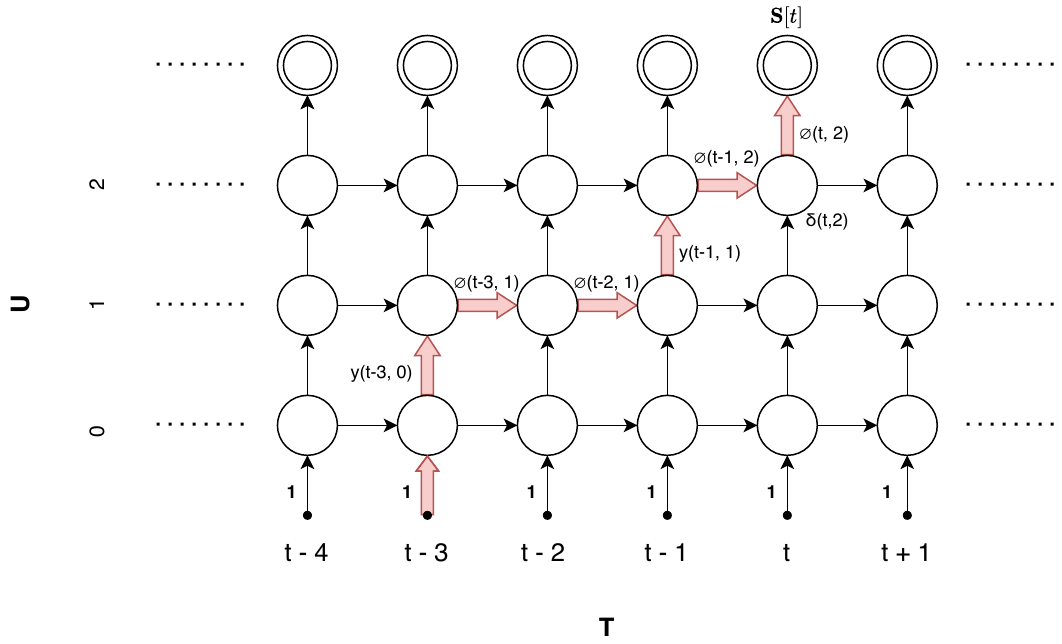}
    \vspace{-1.4em}
    \caption{Decoding path for the RNN-T KWS system. Each node~$(t, u)$ represents the highest score~$\delta(t,u)$ obtained for outputting the first $u$ element of the keyword up to time $t$. The horizontal arrow from node $(t, u)$ indicates the probability $\phi(t, u)$ of outputting a blank, while the vertical arrow represents the probability $y(t, u)$ of emitting the $(u+1)$-th keyword element at time $t$. The optimal keyword path at time $t$ follows the highest-scoring sequence, highlighted by red arrows, corresponding to the most probable keyword occurrence at time $t$.}
    \label{fig:rnnt-kws-decoding-lattice}
\end{figure}

Previous Transducer-based KWS approaches adopt ASR inference strategies, such as greedy search or beam search, without considering the unique characteristics of KWS. Consequently, these methods are suboptimal for frame-level keyword detection. Unlike ASR, where the goal is to generate complete transcriptions, KWS focuses solely on detecting the presence of predefined keywords. To address this, the proposed streaming decoding approach feeds only the keyword token sequence to the predictor, rather than partial hypotheses as in ASR decoding. This design is analogous to RNN-T loss computation and the teacher-forcing strategy used in training autoregressive models. By constraining the decoding space to keyword search, this approach mitigates error accumulation and improves detection accuracy. We denote acoustic feature $\mathbf{x} = \{x_{1},x_{2},...,x_{T}\}$ as before, and keyword transcription~$\mathbf{y} \mkern-2mu = \mkern-2mu \{y_0\mkern-5mu=\mkern-5mu\phi_{\text{RNN-T}}, y_1, y_2, \dots, y_U\}$
($y_0$ denotes the blank symbol). We represent the token/blank emission probability for the decoding lattice following the standard Transducer literature~\cite{icml2012-graves-rnnt} as follows:
\begin{equation}
y(t, u) = p_{t,u}(y_{u}) = P(y_{u+1} \mid \mathbf{x}_{[1\colon t]}, \mathbf{y}_{[0\colon u]}),
\end{equation}
and
\begin{equation}
\phi(t, u) = p_{t,u}(\phi_{\text{RNN-T}}) = P(\phi_{\text{RNN-T}} \mid \mathbf{x}_{[1\colon t]}, \mathbf{y}_{[0\colon u]}),
\end{equation}
for $t \in \left[1, T\right]$ and $u \in \left[0, U\right]$.

For Transducer models, decoding employs a decoding lattice, as illustrated in \Cref{fig:rnnt-kws-decoding-lattice}. We define $\delta(t, u)$ as the highest-scoring path among all paths reaching node $(t, u)$. In streaming decoding, the keyword can begin at any point in the speech stream, requiring special attention to its starting time. To address this, we assign a score of 1 to $\delta(t, 0)$ at each time step $t$, enabling keyword detection from any moment and facilitating seamless wake word detection in continuous speech. Using dynamic programming, as shown in \Cref{alg:alg_rnnt_tdt_algo}, we efficiently compute the complete path score $\delta(t, U)$ for the keyword. The keyword confidence at time $t$ is then obtained by multiplying the path score $\delta(t, U)$ with the blank score $\phi(t, U)$:
\begin{equation}
\mathbf{S}[t] = \delta(t, U) \cdot \phi(t, U),
\end{equation}
where \( \phi(t, U) \) represents the final blank score at time \( t \). Compared to our previous work~\cite{icassp2024-yuxi-tdt_kws}, we introduce three enhancements to the decoding algorithm. First, we incorporate a bonus score \( S_{\text{bonus}} \) to sharpen activation time boundaries, improving keyword endpoint accuracy. Next, we introduce \( T_{\text{out}} \) to filter out excessively long decoding paths. Additionally, we normalize the best path score by its length, converting it into a confidence measure within the range \([0,1]\). This normalization facilitates the fusion process in multi-head decoding. For clarity, we describe \Cref{alg:alg_rnnt_tdt_algo} assuming the input consists of the complete posterior matrix $\mathbf{p} = [p_1, p_2, \dots, p_T] \in \mathbb{R}^{T \times U \times V}$, where $V$ denotes the vocabulary including the blank token. However, the decoding process remains fully streaming with no mandatory latency. In practice, decoding lattice states are maintained, and only one-frame logits or posteriors are fed into the decoding module at each step. Further details are provided in \Cref{alg:alg_rnnt_tdt_algo}.

\subsection{Multi-head Frame-synchronous KWS System}
\label{sec:mfs}

In this section, we describe how to extend Transducer-based KWS system to a multi-head frame-synchronous~(MFS) CTC-Transducer joint framework. Connectionist Temporal Classification~(CTC) is a widely used sequence-to-sequence training criterion for non-autoregressive ASR models. Similar to the RNN-T loss, CTC introduces a special blank token $\phi_{\text{CTC}}$ to model non-speech outputs, facilitating alignment learning between speech and transcriptions. However, its alignment transition rules differ from those of RNN-T. Using the speech input $\mathbf{x}$ and corresponding label $\mathbf{y}$ as defined in \Cref{sec:2.1}, the CTC loss is formulated as:
\begin{align}
\mathcal{L}_{\mathrm{CTC}}\left(\mathbf{x},\mathbf{y}\right)=-\log p\left(\mathbf{y}|\mathbf{x}\right)=-\log \, \sum_{\mathclap{\hspace{0.5em} \mathbf{\pi_\mathrm{CTC}} \in \mathcal{B}_\mathrm{CTC}^{-1}}}\hspace{0.5em}p\left(\bm{\pi}_\mathrm{CTC}|\mathbf{x}\right).
\end{align}
Here, $\mathcal{B}_{\mathrm{CTC}}$ maps valid CTC alignments $\bm{\pi}_{\mathrm{CTC}}$ (including $\phi_{\mathrm{CTC}}$) to the label sequence $\mathbf{y}$, while $\mathcal{B}_{\mathrm{CTC}}^{-1}$ represents its inverse mapping. The CTC-Transducer-based KWS models are trained using a combination of RNN-T and CTC losses, formulated as follows:
\begin{align}
\label{equ:mfs-loss}
\mathcal{L}_\mathrm{MFS}=\mathcal{L}_\mathrm{RNN-T}\left(\mathbf{x},\mathbf{y}\right) + \alpha\mathcal{L}_\mathrm{\mathrm{CTC}}\left(\mathbf{x},\mathbf{y}\right),
\end{align}
where $\alpha$ is a coefficient to control the effect of CTC. In this work, we set $\alpha$ to 0.3 as default. 

CTC assumes conditional independence along the time axis, predicting each frame’s distribution without historical context. While this limits ASR, where contextual information ensures semantic coherence, it may benefit KWS by reducing error accumulation and mitigating false alarms, thereby preventing overfitting to keywords. Moreover, CTC serves as a regularization mechanism in acoustic modeling, improving convergence in multi-task training frameworks. 

For inference, we propose a multi-head frame-synchronous (MFS) decoding algorithm, extending Transducer-based decoding to a joint Transducer-CTC strategy. The CTC branch decoding concept, first introduced in~\cite{icassp2025-yuxi-cdc_kws} and referred to as CTC-Streaming-Decoding in this paper, is a frame-level streaming decoding strategy designed for CTC-based KWS systems. Building on the previously proposed Transducer-based algorithm in~\Cref{alg:alg_rnnt_tdt_algo} and drawing inspiration from the CTC-based approach, we introduce MFS decoding, which performs inference on both branches simultaneously and merges frame-level confidence scores. The inference details are provided in~\Cref{alg:mfs_joint_decoding}. Notably, although the pseudo-code iterates over the frame index \( t \), it does not incur redundant computations in practice. The loop structure is included purely for clarity.

\subsection{Multi-head Frame-asynchronous System}
To enhance performance in challenging environments and reduce computational overhead for faster inference, we extend our MFS KWS to a multi-head frame-asynchronous (MFA) version. Specifically, we modify the frame-level RNN-T and CTC sub-modules into token-level systems. We introduce a token-level variant of the Transducer, Token-and-Duration Transducer~(TDT), which models token duration when predicting token posteriors, replacing the conventional RNN-T. For the CTC branch, we replace frame-synchronous decoding~(FSD) with phone-synchronous decoding~(PSD), which accelerates inference by skipping frames with blank probabilities exceeding a preset threshold. Both approaches can skip irrelevant frames and improve decoding efficiency by reducing unnecessary frame computations.

TDT enhances the conventional Transducer by incorporating token duration prediction into the joiner output. Specifically, while the conventional RNN-T models the posterior distribution $P(v|t,u)$, where $v$ represents a token in the vocabulary or the blank symbol $\phi_{\text{RNN-T}}$ and $t, u$ denote acoustic frame and text token indices, respectively, TDT extends this to a joint distribution $P(v,d|t,u)$, where $d$ represents the predicted token duration at position $(t,u)$. TDT formulates duration prediction as a classification task, where $d$ is selected from $\{0,1,2,\cdots,\mathcal{D}_{\text{max}}\}$, with $\mathcal{D}_{\text{max}}$ being a predefined hyperparameter for duration modeling. In~\cite{icml2023-hainanxu-tdt}, a conditional independence assumption was adopted, leading to the factorization:
\begin{equation}
    P(v,d|t,u) = P_{T}(v|t,u)P_{D}(d|t,u),
\end{equation}
where \( P_{T}(\cdot) \) and \( P_{D}(\cdot) \) represent the token and duration distributions, respectively. Training a CTC-TDT-based KWS system involves replacing the RNN-T loss \( \mathcal{L}_\mathrm{RNN-T} \) in~\Cref{equ:mfs-loss} and RNN-T-based decoding with the TDT loss \( \mathcal{L}_\mathrm{TDT} \) and TDT-based decoding, while keeping all other components unchanged. The MFA loss is defined as:  
\begin{align}
\label{equ:mfa-loss}
    \mathcal{L}_{\mathrm{MFA}} 
    = \mathcal{L}_{\mathrm{TDT}}( \mathbf{x}, \mathbf{y} ) 
    + \alpha \mathcal{L}_{\mathrm{CTC}}( \mathbf{x}, \mathbf{y} )
\end{align}
where details of $\mathcal{L}_\mathrm{TDT}$ can be found in~\cite{icml2023-hainanxu-tdt}.

For the CTC branch, we replace FSD with PSD, which leverages the peaky property of CTC output to filter out blanks which are irrelevant to the key speech frames. The probability of one complete alignment $p(\bm{\pi}|\mathbf{x})$ can be divided into the combination of non-blank frames and blank frames as: 
\begin{align} 
p(\bm{\pi}_{\text{\text{CTC}}}|\mathbf{x}) &= \prod_{t=1}^{T} p(\pi_t|x_t) \label{eq:decoding-non-streaming} 
    = \prod_{\substack{t=1 \\ \text{\textit{non-blank}}}}^{T} p(\pi_t|x_t) \cdot \prod_{\substack{t=1 \\ \textit{blank}}}^{T} p(\pi_t|x_t), 
\end{align}
where {\textit{non-blank} denotes the frame containing acoustic information and \textit{blank} denotes non-speech frame. 
In further, leveraging the peaky phenomenon of CTC, which means the prediction is confident and the maximum probability is almost near to 1 at each time step, we can skip some frames with high probability of blank to accelerate the decoding speed:
\begin{align}
p(\bm{\pi_{\text{\text{CTC}}}}|\mathbf{x})  &= \prod_{\substack{t=1 \\ \text{\textit{non-blank}}}}^{T} p(\pi_t|x_t) \cdot \prod_{\substack{t=1 \\ \text{\textit{blank}}}}^{T} p(\pi_t|x_t) \\
    &= \prod_{\substack{t=1 \\ p(\phi|x_t) < \lambda_{\phi}}}^{T} p(\pi_t|x_t) \cdot \prod_{\substack{t=1 \\ p(\phi|x_t) \geq  \lambda_{\phi}}}^{T} p(\pi_t|x_t) \\ 
    &= \prod_{\substack{t=1 \\ p(\phi|x_t) < \lambda_{\phi}}}^{T} p(\pi_t|x_t) \cdot \prod_{\substack{t=1 \\ p(\phi|x_t) \geq \lambda_{\phi}}}^{T} \underbrace{p(\phi_{\text{\text{CTC}}}|x_t)}_{ \approx 1} \\ 
    &\approx \prod_{\substack{t=1 \\ p(\phi|x_t) < \lambda_{\phi}}}^{T} p(\pi_t|x_t),
\end{align}
where $\lambda_{\phi}$ is a pre-defined threshold to filter frames whose blank probabilities of CTC are very high. The total path can be approximated by those frames contain speech contents with low blank probabilities. This is called phone synchronous decoding~(PSD) for CTC decoding, a technology that can maintain performance and also achieve significant fast decoding speed. 

The threshold of filtering blank frames decides the percentage of frame keeping and it can be tuned to trade off the efficiency and performance.
Compared to each original system~(TDT v.s RNN-T, PSD v.s FSD), TDT and PSD can both gain the ability to model token duration additionally. Combined with the fused strategies proposed in the next section, the joint CTC-Transducer-based KWS system is evolved to a multi-head frame-asynchronous~(MFA) KWS system.

\subsection{Multi-head Joint Decoding}
\label{sec:fusion_strategy}
In our MFA decoding framework, since detection scores from CTC and Transducer are not aligned along the time axis, two critical challenges must be addressed for frame-asynchronous decoding: how to fuse and when to fuse the scores. The fusion method determines the  effectiveness of the system, while the timing of fusion impacts its efficiency. We investigate several different kinds of fused strategies for MFA decoding in this section. The basic idea of the score fusing can be shown as:
\begin{equation}
    \mathbf{S}_{t}^{\mathrm{Fused}} = \mathbf{S}_{t}^{\mathrm{Trans}} \oplus \mathbf{S}_{t}^{\mathrm{CTC}},
\end{equation}
where \( \oplus \) denotes the frame-by-frame fusion operation, and \( \mathbf{S}_{t}^{\mathrm{Trans}} \) and \( \mathbf{S}_{t}^{\mathrm{CTC}} \) represent the confidence scores at time \( t \) from the Transducer and CTC decoding strategies, respectively. To handle frames skipped by TDT or PSD, we introduce Placeholder (PH), a special state representing the decoding score for the skipped frame. The placeholder value is determined by the selected fusion strategy. Since the scores can be either frame-synchronous or -asynchronous, fusion strategies must be carefully designed. We explore several fusion approaches, as illustrated in the right part of \Cref{fig:overview} and detailed below.
\begin{itemize}
    \item \textbf{CTC-Dom~(CTC-Domination)}: The final fused confidence score is determined primarily by the CTC branch. If the CTC branch produces the normal score $\mathbf{S}_{t}^{\mathrm{CTC}}$ at time $t$, it is directly used as the frame-wise confidence. If the CTC branch skips the frame and $\mathbf{S}_{t}^{\mathrm{CTC}}$ is equal to PH, the score is padded by the Transducer branch $\mathbf{S}_{t}^{\mathrm{Trans}}$. If both branches are PHs at time $t$, it implies that both agree the current frame contains no meaningful keyword information. In this case, we assign the confidence score 0 to mark it as a null frame.
    \item \textbf{Transducer-Dom}: The fused confidence score is primarily determined by the Transducer branch, following the inverse logic of CTC-Dom. If the Transducer branch provides a normal score \( \mathbf{S}_{t}^{\mathrm{Trans}} \), it is directly used as the frame-wise confidence. If the Transducer output \( \mathbf{S}_{t}^{\mathrm{Trans}}\) is PH, the score is taken from the CTC branch \( \mathbf{S}_{t}^{\mathrm{CTC}} \). If both branches are PH, the confidence score is set to 0, marking it as a null frame.
    \item \textbf{Equivalence-Dom}. Both branches contribute equally to the final confidence score. If both branches provide decoding scores at frame \( t \), their average is used. If only one branch produces a score, it is directly used. If neither branch are PHs, score 0 is assigned to fill the frame.
\end{itemize}

These fusion strategies operate on single frames, making them straightforward and interpretable. However, they do not account for historical activation states. A high activation score in an isolated frame may indicate a false alarm. Incorporating historical decoding states into the fusion strategy improves reliability and discrimination. {\bf Cross-layer Discrimination Consistency} (CDC), initially proposed in~\cite{icassp2025-yuxi-cdc_kws}, is a score-fusion strategy that exploits the distinct behaviors of intermediate and output layers to enhance performance. We extend CDC by capturing score trend discrimination within a sliding window between CTC and Transducer, using the degree of discrimination as a weighting coefficient for score fusion. At time step $t$, the coefficient $w^\mathrm{CDC}_{t}$ is computed as follows:
\begin{align}
    w^{\mathrm{CDC}}_{t} = f_{\mathrm{sim}}\left( \mathbf{S}^{\mathrm{Trans}}_{[t - \ell_{\mathrm{win}} : t]}, \mathbf{S}^{\mathrm{CTC}}_{[t - \ell_{\mathrm{win}} : t]} \right)
\end{align}
where $\ell_{\mathrm{win}}$ denotes the window size and $f_{\mathrm{sim}}(\cdot)$ is the cosine similarity function. We fix $\ell_{\mathrm{win}} = 20$ across all experiments to ensure fair comparison and reproducibility. Then the fused score is the weighted sum of $\mathbf{S}^{\mathrm{Trans}}_t$ and $\mathbf{S}^{\mathrm{CTC}}_t$:
\begin{align}
    \mathbf{S}_{t}^{\mathrm{Fused}} = (\mathbf{S}_{t}^{\mathrm{Trans}} +
w^\mathrm{CDC}_{t} \cdot \mathbf{S}_{t}^{\mathrm{CTC}})/(1+w^\mathrm{CDC}_{t}).
\end{align}

For MFA decoding, PHs must be replaced with appropriate values before applying CDC-based fusion. We consider two consistency-based strategies:  
\begin{itemize}  
    \item \textbf{CDC-Zero}: In this approach, if a PH is encountered during CDC computation, a score of 0 is assigned, assuming no speech event occurs at that frame.  
    \item \textbf{CDC-Last}: Here, skipped frames are treated as continuations of the most recent state. The nearest preceding non-placeholder value is used for PH replacement.  
\end{itemize}  
We investigate multi-head decoding with various fusion strategies across multiple keyword datasets. The results are presented in~\Cref{sec:fusion-strategies}.

\section{Experimental Setup}
\subsection{Data Configuration}
\label{content:data_configuration}
We evaluate MFA-KWS framework across diverse scenarios:  
1) Fixed single-keyword utterances in both English and Mandarin.  2) Arbitrary keyword detection in continuous speech, covering 20 distinct keywords.  3) Keyword spotting in noisy environments.  To ensure a comprehensive assessment, we utilize multiple widely used datasets, as detailed below.
\begin{itemize}
    \item \textbf{Hey-Snips}\footnote{https://github.com/sonos/keyword-spotting-research-datasets}~\cite{icassp2019-Coucke-second_baseline_of_snips_in_wekws}. The Hey Snips dataset is an open-source KWS dataset that uses ``Hey Snips" as the keyword, pronounced as a single phrase without a pause between the two words. It contains 5,799, 2,484, and 2,529 positive utterances and 44,860, 20,181, and 20,543 negative utterances in the train, development, and test sets, respectively. As complete transcripts for negative utterances are unavailable, these segments are used exclusively for false alarm evaluation, not for training. The false alarm dataset is constructed by aggregating all negative utterances across train, development, and test sets, yielding approximately 97 hours of audio.
    \item \textbf{MobvoiHotwords}\footnote{https://openslr.org/87}~\cite{spl2019-jingyonghou-hixiaowen_nihaowenwen}.
    MobvoiHotwords is a Mandarin KWS corpus featuring two keywords: ``Hi Xiaowen" (Xiaowen) and ``Nihao Wenwen" (Wenwen), along with non-keyword speech. It comprises approximately 262 hours of audio from 287k utterances. The dataset is recorded at varying distances from a smart speaker, with background noise such as household sounds (e.g., music, TV) at different signal-to-noise (SNR) ratios. It includes approximately 21.8k, 3.7k, and 10.6k positive utterances per keyword and 131k, 31.2k, and 52.2k negative utterances across the training, development, and test sets, respectively. The negative test set contains about 63 hours of audio.  
    \item \textbf{LibriSpeech}\footnote{https://openslr.org/12}~\cite{LibriSpeech}. LibriSpeech is a widely used speech corpus containing 960 hours of read speech with corresponding transcripts. In addition to training a phoneme-based acoustic model on LibriSpeech, we select 20 specific words from its two test sets to serve as keywords, simulating arbitrary keyword detection in continuous speech. We refer to this evaluation dataset as \textbf{LibriKWS-20}, which comprises two subsets: test-clean and test-other. Both contain the same 20 keywords, which are listed in~\Cref{table:librikws-20-keywords}. To construct a false alarm dataset, we combine audio samples from the test sets that do not contain the selected keywords, resulting in a total duration of 3 hours separately.
    \item \textbf{AISHELL-2}~\cite{arxiv2018-jiayudu-aishell2}. AISHELL-2 is an industrial-scale ASR corpus consisting of 1000 hours of Mandarin speech. We use AISHELL-2 to train a good Mandarin acoustic seed model for subsequent ``Hi Xiaowen" and ``Nihao Wenwen" KWS experiments. 
    \item \textbf{WHAM!}~\cite{interspeech2019-Gordon-wham}. The WHAM! dataset is a near- and far-field ambient noise corpus recorded in urban environments, featuring sounds such as music, instruments, and background noise from restaurants and bars. It captures diverse real-world acoustic conditions across multiple scenarios. To evaluate model robustness in noisy environments, we mix the test portion of the WHAM! dataset with Hey Snips and LibriKWS-20 datasets at varying SNR levels.
\end{itemize}

\textbf{Data preparation.} For the arbitrary keyword experiment, our goal is to assess the generalization ability of the model to different keywords on general ASR data, so we train models on LibriSpeech-960h without a further fine-tuning stage and evaluate them on LibriKWS-20. For fixed-keyword experiments with Snips or MobvoiHotwords, we first pre-train models on a general ASR dataset, either LibriSpeech-960h or AISHELL-2. The model is then fine-tuned using positive samples from Hey Snips or Xiaowen/Wenwen, combined with an equal amount of non-keyword utterances as common data. For Mandarin MobvoiHotwords, due to its intrinsic challenging acoustic condition, we need to use its own non-keyword dataset as the common data. Since the original MobvoiHotwords does not provide transcriptions of negative datasets, we use the pseudo transcriptions~\footnote{https://www.modelscope.cn/datasets/thuduj12/mobvoi\_kws\_transcription} generated by a powerful Mandarin ASR model~\cite{interspeech2022-zhifugao-paraformer}\footnote{https://huggingface.co/funasr/Paraformer-large}, and then randomly pick some common ASR utterances for fine-tuning.

Additionally, we evaluate performance under various noise conditions using the English Hey Snips and LibriKWS-20 datasets. (Note: MobvoiHotwords is excluded since the original data contains noise, preventing controlled SNR adjustments when adding additional noise.) To simulate noisy conditions, each clean waveform from Snips and LibriSpeech is mixed with a randomly selected noise sample from WHAM!, with the SNR drawn from a uniform distribution from 0 to 20 dB. The noisy negative test set is generated following the same procedure as in training, while the positive test set is simulated at fixed SNR levels. Specifically, all positive test samples from Snips and LibriSpeech are mixed with noise at $\{0, 5, 10, 15, 20\}$ dB.

\begin{table}[t]
  \centering
   \caption{Selected keywords in LibriKWS-20.The keywords for test-clean and test-other two datasets are the same.}
    \label{table:librikws-20-keywords}
  \newcolumntype{S}{>{\small}c}
  \begin{resizebox}{1.0\columnwidth}{!}
  {
    \begin{tabular}{S S S S S}
      \toprule
      ~& ~&\multirow{2}*{\textbf{Keywords}}&~&~ \\
      ~& ~&~&~&~ \\
      \midrule
      almost & anything & behind & captain & children \\
      company & continued & country & everything & hardly \\
      himself & husband & moment & morning & necessary \\
      perhaps & silent & something & therefore & together \\
      \bottomrule
    \end{tabular}%
   }
   \end{resizebox}
\end{table}

\subsection{Training Configuration}

\textbf{Acoustic features.} The acoustic features of the input consist of 40-dimensional log Mel filter bank coefficients (FBank) extracted with a 25ms window length and a 10ms window hop. During training, we apply two kinds of data augmentation strategies. We incorporate online speech perturbation~\cite{speed_perturb} and the warping factors are randomly picked from the set \{0.9, 1.0, 1.1\}. Also, SpecAugment~\cite{specaug} is applied with a maximum frequency mask range of $F=10$ and a maximum time mask range of $T=50$. Specifically, we use two masks of each type for one data sample. We splice five frames from the left and right contexts to construct 440-dimensional features as a contextual acoustic frame and set the frame-skipping parameter to 3, resulting in three times subsampling to reduce computational burden. 

\textbf{Model architecture.}  For the shared encoder, we follow the encoder architecture and setup of hyper-parameters of Tiny Transducer~\cite{Tiny-RNNT}. We use the Deep Feedforward Sequential Memory Network~(DFSMN), a lightweight, effective, and also mainstream architecture for KWS tasks~\cite{ijcai2022-haotong-bifsmn,icassp2022-yuxi-fsmn_kws,icassp2022-yueyue-fsmn_kws,icassp2024-yuxi-tdt_kws,icassp2024-yuxi-cobe,icassp2025-yuxi-ntc_kws,icassp2025-yuxi-cdc_kws} as our speech encoder. The DFSMN-based shared encoder consists of 6 layers, with hidden and projection sizes set to 512 and 320, respectively. The auxiliary encoder for CTC consists of an additional 2 layers of the DFSMN block and a linear projection layer. We utilize the stateless predictor implemented in~\cite{nemo}, with context sizes as two and embedding dimensions as 320. The joiner converts the 320-dimensional encoder and decoder outputs into 256-dimensional representations, which are activated and projected to final outputs. For CTC and RNN-T, the final output consists of 70 monophones, derived from the CMU Pronouncing Dictionary ``cmudict-0.7b"~\cite{cmudict}, or 200 Mandarin syllables from a widely used lexicon\footnote{https://github.com/aishell-foundation/DaCiDian}, along with a special blank token \(\phi_\text{CTC}\) or \(\phi_\text{RNN-T}\). For TDT, the output units additionally include duration options ranging from 0 to the maximum duration \(\mathcal{D}_\text{max}\). Both CTC-RNN-T and CTC-TDT contain approximately 3.3M parameters for both English and Mandarin experiments, making them suitable for on-device KWS applications. The additional parameters introduced by TDT for duration prediction are negligible, accounting for less than 0.1\% of the total model size.

\textbf{Training details.} The local batch size is set to 64, with a maximum of 12,288 frames per batch. We use the AdamW optimizer~\cite{adamw-Ilya-frank-iclr2019} with a maximum learning rate of 1e-3 and 10k warm-up steps. The learning rate is halved if the evaluation loss does not improve. Training is terminated if the loss fails to improve for more than three epochs.

\subsection{Evaluation Details}
\textbf{Baselines.} We conduct comprehensive baseline evaluations to assess the performance of our proposed systems. First, we compare the proposed multi-head decoding framework with various ASR-based decoding strategies, including greedy search and beam search for CTC and RNN-T. Additionally, we compare CTC-Transducer multi-head decoding with single-branch streaming KWS decoding strategies and benchmark our system against end-to-end (E2E) KWS models. Furthermore, we evaluate the frame-asynchronous MFA-KWS against the synchronous MFS-KWS.

\textbf{Evaluation metrics.} We report recall or miss rate at different False Alarm Rates (FAR) or fixed numbers of false alarms. Recall is defined as the ratio of true positives to the total number of positives, while the miss rate is given by \((1 - \text{recall})\). FAR is computed as the ratio of false alarms to the total duration of negative samples. ASR-based decoding strategies for KWS consist of two stages: hypothesis generation and keyword matching. As a result, they cannot inherently control false alarms, making it impractical to set a fixed threshold for recall and FAR. To ensure a fair comparison with ASR-based methods, we also report accuracy, which corresponds to recall at FAR = 0.  For the LibriKWS-20 dataset, we present macro-average metrics across the 20 keywords, including macro-accuracy and macro-recall.

\section{Results And Analysis}
\subsection{Impact of Filtering Threshold for PSD}
\label{sec:4.1-lambda}
This section investigates the optimal threshold $\lambda_{\phi}$ for PSD on the Hey Snips dataset. As shown in \Cref{tab:psd}, increasing the skipped frame ratio gradually reduces the recall, with the degradation rate shifting from slow to rapid. When approximately 35\% of frames are skipped, the performance drop remains acceptable. However, further reduction $\lambda_{\phi}$ leads to a significant decrease. To balance accuracy and computational efficiency, we set $\lambda_{\phi}$ to filter out 35\% of frames for PSD in all subsequent experiments, and $\lambda_{\phi}$ is searched on the corresponding development set.

\begin{table}[h]
    \centering
    \caption{Performance comparison of different skipped frame ratios on the Snips dataset at FAR = 0.02. $\lambda_{\phi}$ is the CTC blank threshold.}
    \label{tab:psd}
  \begin{resizebox}{0.85\columnwidth}{!}
  {
    \begin{tabular}{ccccc|cc}
        \toprule
             \multirow{2}{*}{\textbf{\makecell{Skipped Ratio(\%)}}} & \multirow{2}{*}{\textbf{$\lambda_{\phi}$}} & \multirow{2}{*}{\textbf{Recall@FAR=0.02}} \\
            ~ & ~ & ~ \\
        \midrule
             0 & 1.0 & 98.89 \\
            20 & 0.9999 & 98.54 \\
            30 & 0.9997 & 98.46 \\
            \textbf{35} & \textbf{0.9993} & \textbf{98.10}  \\
            40 & 0.9986 &  97.51 \\
            45 & 0.9971 & 94.35 \\
            50 & 0.9946 & 88.41 \\
        \bottomrule
    \end{tabular}
}\end{resizebox}
    \label{tab:lambda_phi-ablation}
\end{table}

\subsection{Single-branch Training vs. Joint Training}

\begin{table}[b]
    \centering
    \caption{Recall comparison of training systems at various FAR levels on Snips: single-branch training (CTC or Transducer) vs. joint-training (CTC and Transducer). Trans. refers to Transducer. $\oplus$ denotes the multi-task training here. The best results within the same decoding head are \textbf{bold} and the best results in the table are \textbf{\underline{bold and underlined}}.}
    \label{tab:single_joint_training}
    \begin{resizebox}{1.0\columnwidth}{!}{
    \begin{tabular}{c|c|cccc}
        \toprule
            \multirow{2}{*}{\textbf{\makecell{Training  Method}}}&  \multirow{2}{*}{\textbf{\makecell{Decoding \\ Head}}} & \multicolumn{4}{c}{\textbf{Recall(\%)@FARs}} \\
            \cmidrule{3-6}
             & &  0.02 & 0.05 & 0.5 & 1.0 \\
            \midrule
            CTC (CDC-KWS~\cite{icassp2025-yuxi-cdc_kws})  &  \multirow{3}{*}{CTC} & - & \textbf{\underline{99.80}} & - & - \\

            CTC (NTC-KWS~\cite{icassp2025-yuxi-ntc_kws})  &   & - & 98.97 & 99.64 & 99.72 \\
            \multirow{1}{*}{CTC $\oplus$ TDT}  &  & \textbf{98.58} & 99.05 & \textbf{99.84} & \textbf{99.88}\\ 
             \midrule
            Trans. (RNN-T~\cite{icassp2024-yuxi-tdt_kws})  & \multirow{3}{*}{Trans.} & 98.08 & 98.30 & 99.80 & 99.88 \\
            Trans. (TDT~\cite{icassp2024-yuxi-tdt_kws})  &   & 98.56 & 98.85 & 99.80 & 99.84 \\
            \multirow{1}{*}{CTC $\oplus$ TDT}  &  & \textbf{99.57} & \textbf{99.68} & \textbf{\underline{99.96}} & \textbf{\underline{99.96}} \\
            \midrule
           \multirow{1}{*}{CTC $\oplus$ TDT} & MFA & \textbf{\underline{99.60}} & \textbf{\underline{99.80}} & \textbf{\underline{99.96}} &  \textbf{\underline{99.96}}\\
        \bottomrule
    \end{tabular}
    }\end{resizebox}
\end{table}

\begin{table}[b]
    \centering
    \caption{Recall comparison of different fusion strategies for multi-head decoding across three test sets with FA fixed at 2. TDT-4 represents TDT with \(\mathcal{D}_{\text{max}} = 4\). The best average (Avg.) performance in each table block is \textbf{bold}, while the overall best results in the table are \textbf{\underline{bold and underlined}}.}
    \label{tab:joint-method-ablation-different-datasets}
    \begin{resizebox}{1.0\columnwidth}{!} {
    \begin{tabular}{cc|c|ccc|c}
    \toprule
         \multirow{2}{*}{\textbf{Trans.}} & \multirow{2}{*}{\textbf{PSD}} &  \multirow{2}{*}{\textbf{Fusion Strategy}} & \multicolumn{3}{c|}{\textbf{Recall or Macro-recall(\%)}} & \multirow{2}{*}{\textbf{Avg.}} \\
    \cmidrule{4-6}
         && & Snips & test-clean & test-other &  \\
    \midrule
        \multirow{10}{*}{RNN-T} & \multirow{5}{*}{\ding{56}} & CTC-Dom & 98.58 & 98.48 & 89.45 & 95.50 \\
        & & Transducer-Dom & 99.57 & 98.70 & 89.36 & 95.88 \\
        & & Equivalence-Dom & \textbf{\black{99.60}} & \textbf{\black{\uline{99.26}}} & \textbf{\black{91.22}} & \textbf{96.69} \\
        & & CDC-Zero & \textbf{\black{99.60}} & \textbf{\black{\uline{99.26}}} & \textbf{\black{91.22}} & \textbf{96.69}\\
        & & CDC-Last & \textbf{\black{99.60}} & \textbf{\black{\uline{99.26}}} & \textbf{\black{91.22}} & \textbf{96.69}\\
    \cmidrule{2-7}
        & \multirow{5}{*}{\ding{51}} & CTC-Dom & 98.66 & 98.37 & 88.62 & 95.22\\
         && Transducer-Dom & 99.57 & 98.70 & 89.36 & 95.88 \\
         && Equivalence-Dom & \textbf{\black{\uline{99.64}}} & 98.90 & 90.66 & 96.40\\
         & & CDC-Zero & 99.57 & 98.90 & 90.66 & 96.38 \\
         & & CDC-Last& 98.73 & \textbf{\black{\uline{99.26}}} & \textbf{\black{92.27}} & \textbf{96.75} \\
    \midrule
        \multirow{10}{*}{TDT-4} & \multirow{5}{*}{\ding{56}} & CTC-Dom & 98.89 & \textbf{\black{99.21}} & 89.58 & 95.89 \\
        && Transducer-Dom & 98.93 & 98.31 & 90.46 & 95.90 \\
        && Equivalence-Dom & 98.89 & 98.63 & 90.00 & 95.84 \\
        && CDC-Zero & 98.93 & 98.18 & 90.33 & 95.81 \\
        && CDC-Last & \textbf{\black{99.57}} & 98.87 & \textbf{\black{92.90}} & \textbf{97.11} \\
    \cmidrule{2-7}
         & \multirow{5}{*}{\ding{51}} & CTC-Dom & 98.81 & 98.21 & 88.61 & 95.21 \\
        && Transducer-Dom & 98.73 & 98.31 & 91.73 & 96.26 \\
        && Equivalence-Dom & 98.73 & 98.63 & 91.63 & 96.33 \\
        && CDC-Zero & 98.77 & 98.63 & 92.36 & 96.59 \\
        && CDC-Last & \textbf{\black{99.60}} & \textbf{\black{99.11}} & \textbf{\black{\uline{93.63}}} & \textbf{\uline{97.45}} \\
    \bottomrule
    \end{tabular}}
    \end{resizebox}
\end{table}
In this section, we evaluate the effectiveness of Transducer-CTC joint training for the KWS task. As shown in \Cref{tab:single_joint_training}, (CTC \(\oplus\) TDT) represents the joint training approach defined in \Cref{equ:mfa-loss}, while other baselines rely solely on either CTC or Transducer loss. From each block, it is evident that each individual branch of our joint system outperforms single-branch CTC-based and Transducer-based baselines. Compared to our previous Transducer-based work, the key distinction in this table is the inclusion of the CTC branch during training. MFA-KWS consistently outperforms RNN-T/TDT KWS across all FAR conditions. Notably, under strict FAR constraints (FAR=0.02/h or 0.05/h), joint training achieves a \textbf{70.14\%} and \textbf{72.17\%} relative miss rate reduction (recall: 98.56\% vs. 99.57\%, 98.85\% vs. 99.68\%) compared to TDT-only training. When further incorporating multi-head MFA decoding, MFA-KWS attains a \textbf{72.22\%} and \textbf{82.61\%} relative reduction in miss rate (recall: 98.56\% vs. 99.60\%, 98.85\% vs. 99.80\%). 

The results demonstrate that the joint training framework effectively integrates the strengths of both CTC and Transducer, resulting in better convergence and consistently improved performance over single-branch training. Additionally, beyond the benefits of joint training, the three decoding result rows for (CTC $\oplus$ TDT) indicate that multi-head decoding further enhances performance. Therefore, in the following sections, we focus exclusively on evaluating the systems using the joint training paradigm defined in \Cref{equ:mfs-loss} or \Cref{equ:mfa-loss}. We adopt the CDC-Last fusion strategy in this section, as it provides the best performance. A detailed analysis of different fusion strategies is presented in the following section.

\subsection{Comparison of Different Fusion Strategies}
\label{sec:fusion-strategies}
We evaluate various fusion strategies for multi-head streaming decoding in \Cref{tab:joint-method-ablation-different-datasets}, which presents results across multiple English datasets. This analysis specifically focuses on both RNN-T and TDT backbones. 

For RNN-T without PSD, it operates in a frame-synchronous manner and is actually a MFS decoding system, while all the rest systems in \Cref{tab:joint-method-ablation-different-datasets} are MFA decoding systems. For a MFS system, CDC-Zero and CDC-Last yield identical results since no frames are skipped. Compared to naive score fusion strategies~(CTC-Dom, Transducer-Dom, Equivalence-Dom), consistency-based fusion methods (CDC-Zero and CDC-Last) more effectively integrate the Transducer and CTC branches. CDC-based approaches enhance positive scores while suppressing false activations caused by branch-specific misjudgments. Among CDC-based strategies, padding with the nearest previous non-skipped score (CDC-Last) consistently outperforms CDC-Zero and other fusion strategies across multiple datasets. This suggests that preserving the latest decoding state is preferable to resetting it for skipped frames. A likely explanation is that TDT and PSD employ asynchronous skipping mechanisms, making recent scores essential for maintaining detection states in each branch and ensuring accurate merging at each time step. As a result, all subsequent multi-head decoding experiments will use the CDC-Last fusion strategy.

\begin{table}[t]
    \centering
    \caption{Recall performance comparison at FA = 2 between RNN-T and TDT-based joint models with FSD/PSD decoding across Snips, test-clean and test-other test sets. Trans. refers to Transducer. RNN-T denotes the standard RNN-T model, while TDT-\(N\) represents TDT with \(\mathcal{D}_{\text{max}} = N\).}
    \label{tab:rnnt-tdt-fsd-psd}
  \begin{resizebox}{1.0\columnwidth}{!}
  {
    \setlength{\tabcolsep}{3.0mm}{ 
    
    \begin{tabular}{c|cc|ccc|c}
    \toprule
         \multirow{2}{*}{\textbf{\makecell{Multi-head \\ Strategy}}} & \multirow{2}{*}{\textbf{Trans.}} & \multirow{2}{*}{\textbf{PSD}} & \multicolumn{3}{c|}{\textbf{Recall or Macro-recall(\%)}} & \multirow{2}{*}{\textbf{Avg.}} \\
    \cmidrule{4-6} 
         & & &Snips & test-clean & test-other & \\
    \midrule
         MFS & RNN-T & \ding{56}  & \textbf{\black{99.60}} & 99.26 & 91.22 & 96.69 \\
    \midrule
       \multirow{11}{*}{MFA}   & RNN-T & \ding{51} & 98.73 & 99.26 & 92.27 & 96.75 \\
    \cmidrule(l){2-7}    
         & \multirow{2}{*}{TDT-2} & \ding{56}  & 99.09 & 98.60 & 93.24& 96.98 \\
         &  & \ding{51}  & 98.70 & 99.06 & 92.64 & 96.80 \\
    \cmidrule(l){2-7} 
         & \multirow{2}{*}{TDT-4} & \ding{56}  &  99.57 & \black{98.87} & 92.90 & \black{97.11} \\
         &  & \ding{51}  & \textbf\black{{99.60}} & 99.11 & \textbf{\black{93.63}} & \textbf{97.45} \\
    \cmidrule(l){2-7} 
         & \multirow{2}{*}{TDT-6} & \ding{56}  & 99.33 & 98.91 & 92.46 & 96.90 \\
         &  & \ding{51} & 99.05 & \textbf{\black{99.51}} & 92.50 & 97.02 \\
    \cmidrule(l){2-7} 
         & \multirow{2}{*}{TDT-8} & \ding{56}  & 99.17 & 98.91 & 91.72 & 96.60  \\
         &  & \ding{51}  & 98.66 & 98.91 & 91.02 & 96.20  \\
    \cmidrule(l){2-7} 
         & \multirow{2}{*}{TDT-10} & \ding{56}  & 96.99 & 99.06 & 91.60 & 95.88  \\
         &  & \ding{51}  & 97.27 & 99.06 & 91.73 & 96.02\\
    \bottomrule
    \end{tabular}
    }}\end{resizebox} 
    \label{tab:speedup-and-recall-comparison-on-psd-and-tdt}
\end{table}

\subsection{Performance Comparison: MFS vs. MFA}
This section explores optimal model configurations for maximizing performance across three keyword datasets, with results summarized in~\Cref{tab:rnnt-tdt-fsd-psd}. The Transducer branch offers two options: conventional RNN-T or TDT, while the CTC branch supports frame-by-frame FSD or token-by-token PSD decoding. As shown in the first row of~\Cref{tab:rnnt-tdt-fsd-psd}, the system using the original RNN-T with FSD decoding for CTC is classified as MFS-KWS, whereas configurations with TDT or PSD decoding fall under MFA-KWS. The results indicate that TDT consistently outperforms RNN-T, particularly on more challenging datasets, such as the LibriKWS-20 test-other set. Furthermore, optimizing the maximum duration \(\mathcal{D}_{\text{max}}\) in TDT improves keyword detection, with \(\mathcal{D}_{\text{max}} = 4\) yielding the best results across all datasets. Compared to MFS-KWS, MFA-KWS (TDT-4 with PSD) reduces the relative miss rate by 23.0\% on average (2.55\% vs. 3.31\%). For all subsequent experiments, we adopt TDT-4 (TDT with \(\mathcal{D}_{\text{max}} = 4\)) as the MFA-KWS backbone.

\subsection{Performance on Fixed Keyword Benchmarks}

In this section, we further show the performance comparison of proposed MFA-KWS framework and various KWS baselines. We present the results of Snips and MobvoiHotwords in \Cref{tab:kws-snips,tab:kws-wenwen}, respectively. 

\begin{table}[h]
    \centering
    \caption{Recall comparison of the proposed multi-head system with end-to-end KWS systems at various FAR levels on the Hey-Snips dataset. Trans. denotes Transducer. Pos. Snips denotes the positive part of Snips. Equ. LS denotes equivalent numbers of LibriSpeech utterances.}
    \label{tab:kws-snips}
  \begin{resizebox}{0.95\columnwidth}{!}
  {
    \begin{tabular}{c|c|ccc}
    \toprule
        \multirow{2}{*}{\textbf{System}} & \multirow{2}{*}{\textbf{\makecell{Training \\ Data}}}  & \multicolumn{3}{c}{\textbf{Recall(\%)@FARs}} \\
    \cmidrule{3-5}
       & & 0.05 & 0.5 & 1.0 \\ 
    \midrule
        RIL-KWS~\cite{is2020-kunzhang-first_baseline_of_snips_in_wekws} & \multirow{3}{*}{\makecell{Official Snips}}  & - & 96.47 & 97.18 \\
        WaveNet~\cite{icassp2019-Coucke-second_baseline_of_snips_in_wekws} &   & - & 99.88 & -\\
        MDTC\cite{icassp2023-jiewang-wekws} &  & - & 99.88 & 99.92 \\
    \midrule
        MDTC & \multirow{2}{*}{\makecell{Pos. Snips \\ + Equ. LS}}  & 89.52 & 98.85 & 99.29 \\
    \cmidrule{1-1} \cmidrule{3-5}
        MFA-KWS &   & \textbf{99.80} & \textbf{99.96} &  \textbf{99.96} \\
    \bottomrule
    \end{tabular} 
    }\end{resizebox} 
\end{table}

\begin{table}[t]
    \centering
    \caption{Recall comparison at FAR = 0.5/h for the Mandarin dataset MobvoiHotwords.}
    \label{tab:kws-wenwen}
  \begin{resizebox}{0.95\columnwidth}{!}
  {
    \begin{tabular}{c|cc|c}
    \toprule
        \multirow{2}{*}{\textbf{System}} & \multicolumn{2}{c|}{\textbf{Recall(\%)}} & \multirow{2}{*}{\textbf{Avg.}} \\
        \cmidrule(lr){2-3} 
        & Xiaowen & Wenwen &\\
    \midrule
        Streaming Transformer~\cite{icassp2021-yimingwang-baseline_of_mobvoi_in_wekws} & 99.30 & 99.50 & 99.40\\
        MDTC~\cite{icassp2023-jiewang-wekws} & 99.50 & 99.57 & 99.54\\
        LF-MMI~\cite{interspeech2020-yimingwang-baseline_of_mobvoi_in_lf_mmi} & 99.60 & 99.50 & 99.55 \\
        Res2Net-KWS~\cite{arxiv2022-yuqiuchen-baseline_of_mobvoi_in-res2net} & 99.53 & 99.75 & 99.64\\
    \midrule
        MFA-KWS & \textbf{99.66} & \textbf{99.92} & \textbf{99.79} \\
    \bottomrule
    \end{tabular}
} \end{resizebox}
\end{table}

\begin{figure*}[t]
    \centering
    \includegraphics[width=1.0\linewidth]{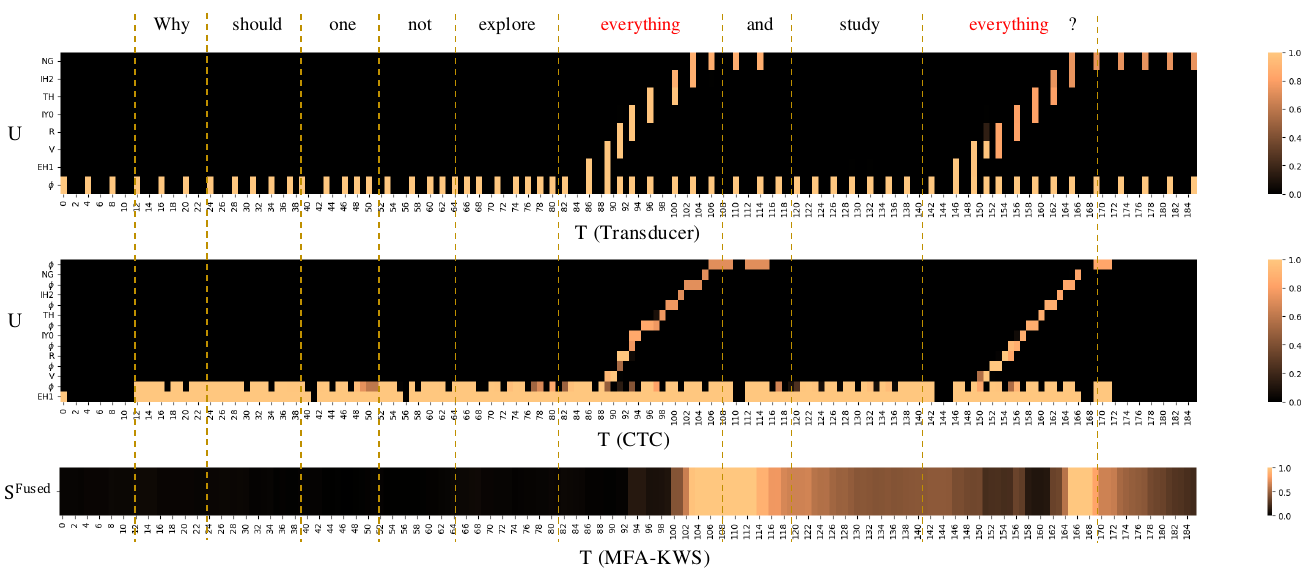}
    \caption{Heatmaps of the wake-up scores at each $(t, u)$ for the Transducer and CTC branches, along with the MFA joint decoding score. The utterance is selected from the test-clean, with {\it everything} as the keyword. The vertical yellow dashed lines indicate word boundaries derived from force alignments.}
    \label{fig:heatmap}
\end{figure*}

For fixed English Keyword dataset Snips, except the comparison on ASR-based systems in~\Cref{tab:single_joint_training}, we also compare with end-to-end KWS systems. To build a bridge between E2E KWS and ASR-based KWS systems, we initially re-implement the best E2E baseline MDTC trained on positive Snips and equivalent LibriSpeech~(denoted as Pos. Snips + Equ. LS), and tested on 97 hours negative test data as mentioned in~\Cref{content:data_configuration}. Compared the results between row 3 and row 4 in~\Cref{tab:kws-snips}, MDTC can still achieve comparable performance under new data strategy (99.88\% vs. 98.85\%; 99.92\% vs. 99.29\%).  It indicates the new dataset construction affect the performance slightly and even makes the dataset more challenging. Then we compare our MFA-KWS with E2E KWS. Although E2E models, such as RIL-KWS, WaveNet and MDTC can achieve good performance under loose FAR test conditions, the proposed MFA-KWS can even detect almost all correctly (99.96\% vs. 99.88\% or 99.92\% under FAR=0.5/h or FAR=1.0/h). It shows the powerful keyword detection capability of MFA-KWS. Meanwhile, as all systems perform well under these FAR conditions, we finally evaluate the performance on extremely strict conditions (FAR=0.02/h or FAR=0.05/h), which is more applicable in realistic situation. We find the recall 89.52\% of MDTC under FAR=0.05 in row 4 is obviously lower than the result of MFA-KWS in the same test conditions, which demonstrates E2E KWS systems cannot process difficult test scenarios well. Moreover, compared to various systems, including CTC-based~\cite{icassp2025-yuxi-cdc_kws}, WFST-based~\cite{icassp2025-yuxi-ntc_kws}, and Transducer-based~\cite{icassp2024-yuxi-tdt_kws} systems in~\Cref{tab:single_joint_training}, and E2E systems~\cite{is2020-kunzhang-first_baseline_of_snips_in_wekws,icassp2019-Coucke-second_baseline_of_snips_in_wekws,icassp2023-jiewang-wekws} in~\Cref{tab:kws-snips}, MFA-KWS consistently achieves superior performance across nearly all FAR conditions, setting a new \textbf{state-of-the-art (SOTA)} benchmark on the Snips dataset.

\Cref{tab:kws-wenwen} further evaluates MFA-KWS on the Mandarin benchmark MobvoiHotwords. While all systems achieve relatively high recall at FAR=0.5/h on Xiaowen and Wenwen, the MFA strategy consistently outperforms them, demonstrating its effectiveness. Compared to the best-performing models on the two datasets, MFA-KWS achieves \textbf{27.66\%} and \textbf{68\%} relative reductions in miss rate (recall: 99.53\% vs. 99.66\%; 99.75\% vs. 99.92\%). These SOTA results confirm that MFA-KWS generalizes well to Mandarin, highlighting its robustness across different modeling units and languages.

\begin{table}[h]
    \centering
    \caption{Performance comparison of MFA-KWS with ASR-based CTC and Transducer KWS baselines. Macro-accuracy is provided for 20 keywords on the test-clean and test-other subsets of LibriKWS-20.}
    \label{tab:decoding-algos-comparison}
  \begin{resizebox}{1.0\columnwidth}{!}
  {
    \begin{tabular}{c|c|cc|c}
        \toprule
        \multirow{2}{*}{\textbf{System}} & \multirow{2}{*}{\textbf{Decoding Algo.}} & \multicolumn{2}{c|}{\textbf{Macro-accuracy(\%)}} & \multirow{2}{*}{\textbf{Avg.}} \\
    \cmidrule{3-4}
         & & test-clean & test-other & ~\\
    \midrule
         \multirow{3}{*}{CTC} & Greedy  & 89.68 & 64.64 & 77.16 \\
          & Prefix Beam Search (beam=10)  & 72.47 & 43.52 & 57.60 \\
          & CTC Streaming~\cite{icassp2025-yuxi-cdc_kws}  & 93.86 & 70.99 & 82.43 \\
    \midrule
         \multirow{3}{*}{Transducer} & Greedy & 83.20 & 51.93 & 67.57\\
          & Beam Search (beam=10)  & 86.30 & 54.83 & 70.57\\
          & Transducer Streaming~\cite{icassp2024-yuxi-tdt_kws}  & 93.18 & 69.71 & 81.45\\
    \midrule
         \multirow{1}{*}{Joint} & MFA & \textbf{94.25} & \textbf{72.75} & \textbf{83.50} \\ 
    \bottomrule
    \end{tabular}
    }\end{resizebox} 
\end{table}

\subsection{Comparison on Arbitrary Keyword Detection}
\label{sec:kws-compare-asr-results}

In this section, we evaluate performance on the arbitrary KWS dataset LibriKWS-20. To enable arbitrary keyword detection, all models are trained on the general ASR dataset LibriSpeech. End-to-end KWS systems are unsuitable for this task as they require keyword-specific data for training. Therefore, we compare the proposed MFA-KWS with various ASR-based KWS systems, incorporating different architectures (CTC and Transducer) and decoding algorithms, including greedy search, beam search, and streaming KWS decoding. Since selecting an optimal operating point for ASR decoding is challenging, we present average accuracy at FAR=0 for fair comparison\footnote{We also compare macro-recall under FAR\(\neq\)0 with the results from Transducer streaming decoding in~\cite{icassp2024-yuxi-tdt_kws}. MFA-KWS outperforms TDT-KWS, and the conclusions remain consistent.} . As shown in \Cref{tab:decoding-algos-comparison}, MFA-KWS achieves the highest accuracy across both test datasets and exhibits the best average accuracy among all decoding strategies.  

Overall, the proposed streaming MFA-KWS not only outperforms ASR-based decoding strategies but also offers flexibility in adjusting the operating point for KWS tasks. Additionally, MFA-KWS surpasses single-branch streaming decoding at FAR=0, confirming its ability to effectively integrate the strengths of both branches for enhanced performance. For FAR\(\neq\)0, we present further arbitrary-keyword results in the next noise-robustness analysis~\Cref{sec:noise-robu}. The best results further highlight the strong capability of MFA-KWS in arbitrary keyword detection.

We present an inference example from LibriKWS-20 to demonstrate MFA decoding in a continuous speech stream in \Cref{fig:heatmap}. The figure includes TDT-based streaming decoding, CTC-based streaming decoding, and the fused detection score derived from CDC-Last at each frame \( t \). It clearly visualizes the decoding states of each node in the decoding lattice, along with the start and end times of activation events.


\subsection{Performance Analysis under Noisy Environments}
\label{sec:noise-robu}
This section further evaluates the robustness of MFA-KWS to noise. \Cref{tab:snips-ls20-snr} reports the performance of different decoding strategies under varying signal-to-noise ratios (SNRs) for Snips and LibriKWS-20. In this table, CTC-FSD and CTC-PSD indicate whether PSD is enabled in CTC-based decoding. The results confirm that the proposed MFA framework consistently outperforms all other methods across the three test sets, demonstrating strong robustness in noisy environments. Several key observations include: 1) Transducer decoding demonstrates greater stability and reliability on continuous speech datasets, particularly on the more challenging test-other set, compared to CTC-based decoding. 2) Under noisy conditions, frame-skipping strategies consistently enhance performance (MFA vs. MFS, CTC-PSD vs. CTC-FSD), likely by reducing noise interference and improving focus on critical speech frames. The frame-asynchronous approach may help the model prioritize essential speech-content frames while filtering out irrelevant or noisy frames. 3) Multi-head decoding significantly outperforms single-branch methods in noisy environments, with MFA-KWS maintaining stable effectiveness across various SNR levels. These results highlight the strong noise robustness of MFA-KWS, demonstrating its ability to handle complex acoustic scenarios effectively.

\begin{table}[t]
    \caption{Performance comparison of different decoding strategies across Snips, test-clean, and test-other datasets at fixed FA = 2 under varying noise conditions. MFS is trained using a combination of CTC and RNN-T, while others are trained on CTC and TDT with \(\mathcal{D}_\mathrm{max} = 4\).}
    \centering
    \label{tab:snips-ls20-snr}
  \begin{resizebox}{1.0\columnwidth}{!}
  {
    \begin{tabular}{cc|ccc|ccccc|c}
    \toprule
      \multirow{2}{*}{\textbf{Dataset}} & \multirow{2}{*}{\textbf{\makecell{Decoding \\ Algo.}}} &  \multirow{2}{*}{\textbf{Trans.}} & \multirow{2}{*}{\textbf{CTC}} & \multirow{2}{*}{\textbf{PSD}} & \multicolumn{5}{c|}{\textbf{Recall or Macro-recall(\%)@SNRs}} & \multirow{2}{*}{\textbf{Avg.}} \\
    \cmidrule{6-10}
         & & & &  & 0 & 5 & 10 & 15 & 20  & \\
    \midrule
        \multirow{5}{*}{Snips}     & CTC-FSD &  \ding{56} & \ding{51} & \ding{56} &  66.47 & 84.18 & 91.02 & 94.58 & 95.93 & 86.44  \\
        & CTC-PSD &  \ding{56} & \ding{51} & \ding{51} & 76.35 & 89.13 & 93.83 & 95.97 & 96.80 & 90.42  \\
        & TDT & \ding{51} & \ding{56} & \ding{56} &  66.51 & 84.26 & 91.06 & 94.66 & 96.01 & 86.50  \\
         & MFS & \ding{51} & \ding{51} & \ding{56} &  78.96 & 92.17 & 96.99 & \textbf{\black{98.46}} & 98.70 & 93.06  \\
        & MFA & \ding{51} & \ding{51} &\ding{51} & \textbf{\black{83.35}} & \textbf{\black{93.16}} & \textbf{\black{97.27}} & 98.42 & \textbf{\black{98.89}} & \textbf{94.22}  \\
    \midrule
        \multirow{5}{*}{test-clean}  & CTC-FSD & \ding{56} & \ding{51} &\ding{51} & 73.66 & 92.61 & 97.75 & \textbf{\black{99.28}} & 98.68  & 92.40 \\
        & CTC-PSD & \ding{56} & \ding{51} & \ding{56}& 75.74 & 92.95 & 97.61 & 98.89 & 98.06 & 92.65  \\
           & TDT & \ding{51} & \ding{56} & \ding{56} & 77.83 & 91.62 & 97.14 & 98.01 & 98.82 & 92.68\\     
        & MFS & \ding{51} & \ding{51} & \ding{56} & 77.61 & 92.04 & 98.55 & 98.79 & \textbf{\black{98.98}} & 93.19  \\
        & MFA & \ding{51} & \ding{51} &\ding{51} & \textbf{\black{79.82}} & \textbf{\black{95.07}} & \textbf{\black{98.84}} & \textbf{\black{99.28}} & 98.82 & \textbf{94.37}  \\
    \midrule
        \multirow{5}{*}{test-other} & CTC-FSD & \ding{56} & \ding{51} &\ding{51} & 42.30 & 66.52 & 76.46 & 81.75 & 84.81 & 70.37 \\
        & CTC-PSD & \ding{56} & \ding{51} & \ding{56}& 43.16 & 67.15 & 76.35 & 82.12 & 85.54 & 70.86  \\ 
        & TDT & \ding{51} & \ding{56} & \ding{56} & 50.15 & 70.29 & 80.49 & \textbf{\black{84.83}} & 87.24 & 74.60  \\    
        & MFS &\ding{51} & \ding{51} & \ding{56} & 44.37 & 70.25 & 80.41 & 84.28 & 85.18 & 72.90  \\
        & MFA & \ding{51} & \ding{51} &\ding{51} &  \textbf{\black{50.16}} & \textbf{\black{71.27}} & \textbf{\black{81.81}} & 84.38 & \textbf{\black{88.15}} & \textbf{75.15}  \\
    \bottomrule
    \end{tabular}
    }\end{resizebox} 
    \label{tab:robustness-ls20}
\end{table}


\subsection{Inference Efficiency}

This section evaluates the efficiency of the proposed MFA-KWS by measuring the relative speed-up across all datasets. As shown in \Cref{tab:speed-up}, MFA-KWS achieves a 1.47×–1.63× speed-up over the base MFS-KWS. It is faster than single-branch RNN-T and comparable in speed to TDT-4, with only CTC-PSD being faster. Given that MFA-KWS significantly outperforms all single-branch systems, these results confirm that the proposed framework not only maintains strong performance but also delivers substantial efficiency improvements.

\begin{table}[h]
    \centering
    \caption{Relative speed-up comparison of all decoding algorithms across all test sets, with MFS decoding speed as the benchmark.}
    \label{tab:speed-up}
  \begin{resizebox}{1.0\columnwidth}{!}
  {
    \begin{tabular}{c|ccccc}
    \toprule
         \multirow{2}{*}{\textbf{Decoding Algo.}} & \multicolumn{5}{c}{\textbf{Relative Speed-up}} \\
    \cmidrule{2-6} 
         & Snips & test-clean & test-other & Xiaowen & Wenwen  \\
    \midrule
           MFS & 1.0X & 1.0X & 1.0X & 1.0X & 1.0X \\
 \midrule
 
           CTC-PSD & 2.03X & 1.87X & 1.77X & 1.93X & 2.09X \\
 
           RNN-T & 1.19X & 1.33X & 1.10X & 1.26X & 1.23X \\

           TDT-4 & 1.64X & 1.59X & 1.60X & 1.57X & 1.62X \\
    \midrule
           MFA & 1.63X & 1.55X & 1.53X & 1.52X & 1.47X \\
    \bottomrule
    \end{tabular}
    }\end{resizebox} 
    \label{tab:speedup-and-recall-comparison-on-psd-and-tdt}
\end{table}

\section{Conclusions}

In this work, we propose MFA-KWS, a joint CTC-Transducer KWS system with keyword-specific multi-head frame-asynchronous decoding. MFA-KWS mitigates error accumulation in Transducer-based KWS and significantly improves discrimination between keywords and challenging negative samples. Additionally, we explore several fusion strategies for multi-head frame-asynchronous decoding, with CDC-Last consistently delivering robust performance across various test conditions. Compared to multiple baselines, MFA-KWS demonstrates superior performance across different datasets and noise conditions. Moreover, the MFA framework achieves a 47\%–63\% speed-up over MFS-KWS. Overall, our results establish MFA-KWS as an effective and efficient KWS framework, well-suited for on-device deployment.

\bibliographystyle{IEEEtran}
\bibliography{citations/refs.bib}

\end{document}